\documentclass[twocolumn,showpacs,preprintnumbers,amsmath,amssymb]{revtex4}

\usepackage{graphicx}
\usepackage{dcolumn}
\usepackage{bm}

\begin{document}

\preprint{APS/123-QED}

\title{NMR and NQR study of pressure-induced superconductivity and the origin of critical-temperature enhancement in the spin-ladder cuprate Sr$_2$Ca$_{12}$Cu$_{24}$O$_{41}$ }

\author{N. Fujiwara, Y. Fujimaki$^1$, S. Uchida$^1$, K. Matsubayashi$^2$, T. Matsumoto $^{2, 3}$ and  Y. Uwatoko$^2$  }

\affiliation{ Graduate School of Human and Environmental Studies,
Kyoto University, Yoshida-Nihonmatsu-cyo, Sakyo-ku, Kyoto 606-8501,
Japan}

\altaffiliation{ Email: naokif@mbox.kudpc.kyoto-u.ac.jp}

\affiliation{$^1$Department of Superconductivity, University of
Tokyo,7-3-1 Hongo, Bunkyo-ku, Tokyo 113-8656, Japan }

\affiliation {$^2$Institute for Solid State Physics, University of
Tokyo, 5-1-5 Kashiwanoha, Kashiwa, Chiba 277-8581, Japan}

\affiliation{$^3$National Institute for Materials Science, Tsukuba
305-0047, Japan}

\date{06 August 2009}

\begin{abstract}
Pressure-induced superconductivity was studied for a spin-ladder
cuprate Sr$_2$Ca$_{12}$Cu$_{24}$O$_{41}$ using nuclear magnetic
resonance (NMR) under pressures up to the optimal pressure 3.8 GPa.
Pressure application leads to a transitional change from a
spin-gapped state to a Fermi-liquid state at temperatures higher
than $T_c$. The relaxation rate $1/T_1$ shows activated-type
behavior at an onset pressure, whereas Korringa-like behavior
becomes predominant at the optimal pressure, suggesting that an
increase in the density of states (DOS) at the Fermi energy leads to
enhancement of $T_c$. Nuclear quadrupole resonance (NQR) spectra
suggest that pressure application causes transfer of holes from the
chain to the ladder sites. The transfer of holes increases DOS below
the optimal pressure. A dome-shaped $T_c$ versus pressure curve
arises from naive balance between the transfer of holes and
broadening of the band width.

\end{abstract}

\maketitle

 A spin-ladder system, Sr$_{14-x}$Ca$_{x}$Cu$_{24}$O$_{41}$ (x=11.5-13.5),
 which is known as the only cuprate superconductor with non-square lattice Cu-O layers,
 offers a good model for study of quasi-one-dimensional
 superconductivity. To date, various theoretical studies have been performed for isolated two-leg ladders
 or the Trellis lattice. $^{1-11}$
Although the system is attractive for comparison with
high-\emph{$T_c$} cuprates, little experimental effort has been
devoted to investigation of superconductivity because of
experimental difficulties under high pressure;
 the onset and optimal pressures are around 3 and 4 GPa, respectively. $^{12, 13}$
 The experiments at high pressures, above 3 GPa, have been performed
 only for resistivity, AC susceptibility, X-ray diffraction and NMR measurements. $^{2-21}$
 Appearance of a dome-shaped \emph{$T_c$} curve on the
 $P-T$ phase diagram and a decrease in the spin gap caused by
 pressure application
are reminiscent of hole doping in high-\emph{$T_c$} cuprates. $^{13,
17-18}$ The pressure-induced superconducting state survives even at
high fields beyond the Pauli paramagnetic limit, when the field
$\textbf{\emph{H}}$ is applied parallel to
  the leg direction ($\mathbf{H} \parallel$ c), for which orbital suppression is depressed. $^{15}$
  The NMR measurement at 3.5 GPa shows that the quasi-particle excitation accompanies a full
  gap;
  however, the Knight shift shows no appreciable change below \emph{$T_c$}. $^{20}$

This system also shows features common to quasi-one-dimensional
systems such as organic superconductors. An example is emergence of
a charge density wave (CDW) in a lightly doped regime $(x\leq 0.8)$
at ambient pressure. $^{22, 23}$ The stable superconducting state
beyond the Pauli paramagnetic limit and no change of the Knight
shift below $T_c$
 are also common to (TMTSF)$_2$X (X=ClO$_4$ or PF$_6$). These features have often been discussed
  in relation to the possibility of p-wave superconductivity
 or singlet-Fulde-Ferrell-Larkin-Ovchinnikov (FFLO) phase transition. $^{24-26}$ Recently, in
the cuprate spin-ladder system, the possibility of FFLO state has
been investigated theoretically at high field. $^{8-10}$ NMR study
at optimal pressure is required to investigate the reasons that lead
to $T_c$ enhancement after pressure application. Questions that need
to be addressed are : (1) what is the origin of the dome-shaped
$T_c$ curve, and (2) what is the spin susceptibility ($\chi_{spin}$)
like around \emph{$T_c$}. We performed $^{63}$Cu-NMR measurements up
to the optimal pressure ($\sim$ 3.8 GPa) to study these problems.

NMR and NQR measurements were performed using a single crystal of
Sr$_2$Ca$_{12}$Cu$_{24}$O$_{41}$ having a volume of 4x2x1 mm$^3$.
The measurements up to 3.5 GPa were carried out using a clamped-type
pressure cell.  A special variable-temperature insert (VTI) equipped
 with  a 45 tons press was used to achieve a high pressure of 3.8 GPa.
 Details of the apparatus are described in Fujiwara $\emph{et. al.}.$ $^{27}$
 We also determined  \emph{$T_c$} from resistivity measurements using a cubic-anvil cell.
A top-loading insert with a 250 tons press was used to control a
steady load at low temperatures. Details of the apparatus and
calibration method are described in Mori $\emph{et. al.}.$ $^{28}$

Figure 1 shows $1/T_1$ of $^{63}$Cu nuclei measured at 6.2T for
$\emph{\textbf{H}}\parallel$ c up to 3.8GPa. The rate $1/T_1$ shows
activated-type $T$ dependence, which originates from the spin-gap
excitation, followed by $\emph{T}$-linear dependence;
\begin{equation}
 T_1^{-1}  =  A e^ {-\Delta /T } + BT.
\end{equation}
The superconducting state appears at low temperatures accompanied by
a hump, implying existence of a full gap for the quasi-particle
excitation. The coefficient of the $T$-linear dependence [B in Eq.
(1)] is enhanced with increasing pressure up to 3.8 GPa, similar to
\emph{$T_c$}. The enhancement of \emph{$T_c$} is confirmed by a
shift of the coherence peak to a higher temperature. The $1/T_1T$
vs. $T$ plot normalized by those quantities at $T_c$ fits a single
curve. The $T$-linear dependence comes from the Korringa relation,
and thus, the coefficient $B$ is proportional to the square of the
density of states (DOS) at the Fermi energy, $D(E_F)$. Therefore,
the enhancement of $T_c$ is attributable to an increase in $D(E_F)$.
On the other hand, the spin gap $\Delta$ and the coefficient A in
Eq. (1) decrease with increasing pressure. The values of A at
ambient pressure, 3.5 and 3.8GPa are in a ratio of about 6:4:1. The
activated-type contribution fades out with increasing pressure,
although both mechanisms coexist in the relevant pressure regime.
The rate \emph{$1/T_1$} under extremely high pressure is expected to
become $1/T_1T=$constant. The $P$ dependence of both $B$ and
$\Delta$ is shown in Fig. 2, together with that of \emph{$T_c$}
determined from the resistivity measurements. The data determined
from the onset and zero resistivity are plotted in Fig. 2. We also
plotted the data of $T_c$ determined from the inductance of the NMR
probe.
 Our results show that $\Delta$ is almost the same
up to the onset pressure, but decreases drastically above it.
Corresponding to $\Delta$, $B$
 shows a remarkable upturn toward the optimal pressure. The
 upturn of
$B$ (= $T$ independent) means that antiferromagnetic spin
fluctuation hardly affects
 the enhancement of $T_c$; when antiferromagnetic fluctuations are
 predominant like high-$T_c$ cuprates, $T$ dependence of $1/T_1T$
 shows a upturn toward \emph{$T_c$}.

\begin{figure}
\includegraphics{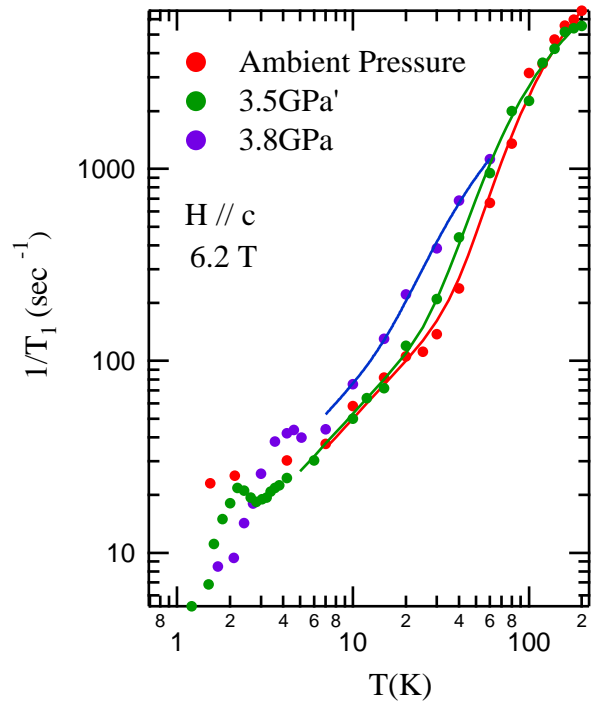}
\caption{\label{fig:epsart} (Color online) Nuclear magnetic
relaxation rate $1/T_1$ of $^{63}$Cu nuclei measured at 6.2 T for
$\textbf{H} \parallel$ c, the leg direction. The data are fitted by
Eq. (1).}
\end{figure}

\begin{figure}
\includegraphics{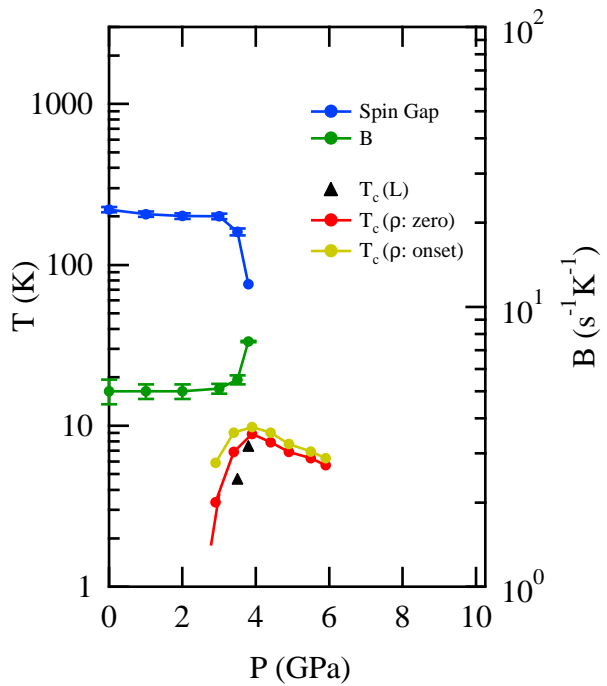}
\caption{\label{fig:epsart} (Color online) Pressure dependence of
spin gap $\Delta$ and the coefficient of $T$-linear term, $B$ in Eq.
(1). Closed triangles represent $T_c$ determined from the inductance
of the NMR probe. $T_c(\rho: onset)$ and $T_c(\rho: zero)$ represent
values of $T_c$ determined from the onset and zero resistivity,
respectively.}
\end{figure}

An increase in $D(E_F)$ is also observed from
 the Knight shift for
$\emph{\textbf{H}} \parallel$ c; The spin-gap behavior observed
clearly at 3.5 GPa almost vanishes at 3.8GPa (See Fig. 3). The
Knight shift is decomposed into the orbital and spin parts;
$K^c=K_{orb}^c + K_{spin}^c$. The orbital part $K_{orb}^c$ is almost
unchanged for a pressure of 3.5-3.8 GPa as explained in the
following paragraphs. Therefore, the spin part $K_{spin}^c$ or
$\chi_{spin}$, which is proportional to $D(E_F)$ in a normal metal,
is enhanced with increasing pressure. We estimated $K_{orb}^c$ from
the NQR measurements because $K_{orb}^c$ is related to the NQR
frequency $\nu$ as
\begin{equation}
K_{orb}^c =\alpha\nu + \beta.
\end{equation} The relation is derived from Eqs. (3-5) in the following paragraphs.

Before presenting the results, we state the NQR results measured at
4.2 K. The NQR spectra at 3.5 GPa and ambient pressure are shown in
Fig. 4. We measured the NQR signals up to 40 MHz with a resolution
of 0.1 MHz/point. The signals originating from the ladder and chain
sites were observed at 10-25 and 30-35 MHz, respectively. At ambient
pressure, $^{63}$Cu and $^{65}$Cu signals are clearly assigned only
in case of chain sites, whereas they overlap with each other in case
of ladder sites. The overlap is decomposed into two gaussian
functions. The resonance position is almost the same as that
measured in an earlier study. $^{29}$ At 3.5 GPa, the signals
arising from the ladder sites move to high frequencies, whereas
those from the chain sites move to low frequencies. The broadening
of the line width is caused by stress anisotropy, which appears when
pressure mediation liquid freezes during the pressurizing process.
The $\emph{P}$ dependence of the NQR frequencies is shown in Fig. 5.
The $^{63}$Cu signals move by 2.27 and -0.69 MHz for the ladder and
chain sites, respectively, and those of $^{65}$Cu move by 1.05 and
-0.60 MHz for the ladder and chain sites, respectively. The $\nu$ in
Eq. (2) is determined by two factors; the electric field gradient
(EFG) arising from the hole number of the $3d_{x^2-y^2}$ orbital,
$n_{3d}$ and that arising from the surrounding 4p orbitals. In
high-\emph{$T_c$} cuprates, the latter contribution is not so
sensitive to doping level, therefore $\nu$ is determined mainly by
$n_{3d}$ ;
\begin{figure}
\includegraphics{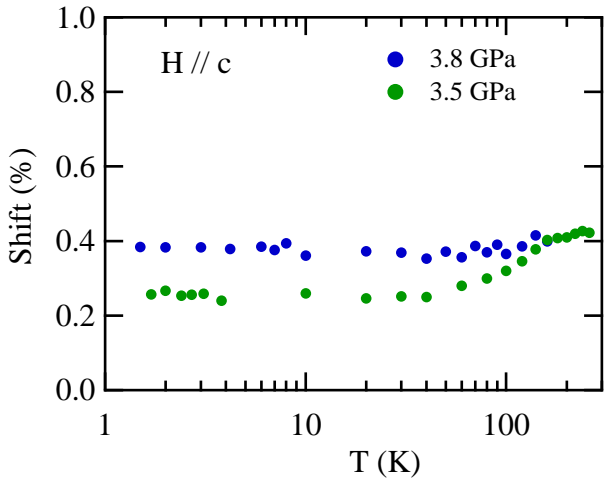}
\caption{\label{fig:epsart}(Color online) Knight shift of $^{63}$Cu
nuclei for the ladder sites at low temperatures derived from high
fields.}
\end{figure}
\begin{equation}
\nu=n_{3d} \nu_{3d} + \nu_{4p}
\end{equation}
where $\nu_{4p} \simeq$ -65 MHz and $\nu_{3d} \simeq$ -117 MHz are
estimated. $^{30}$ Equation (3) can be applied to both ladder and
chain sites of Sr$_{14-x}$Ca$_{x}$Cu$_{24}$O$_{41}$ as well as
high-$T_c$ cuprates. The increase in the number of holes at the
ladder Cu sites comes from the chain sites. We estimated $n_{3d}$
from the resonance frequencies of $^{63}$Cu as 0.716 and 0.736 at
ambient pressure and 3.5 GPa, respectively.
\begin{figure}
\includegraphics{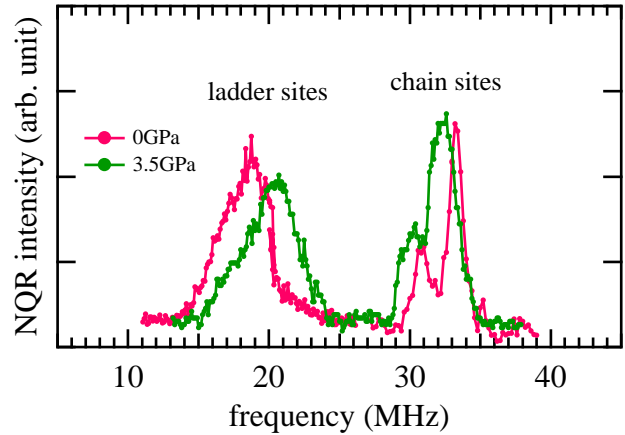}
\caption{\label{fig:epsart}(Color online) NQR spectra measured at
4.2K for the ladder (15-25MHz) and chain sites (30-35 MHz).}
\end{figure}

\begin{figure}
\includegraphics{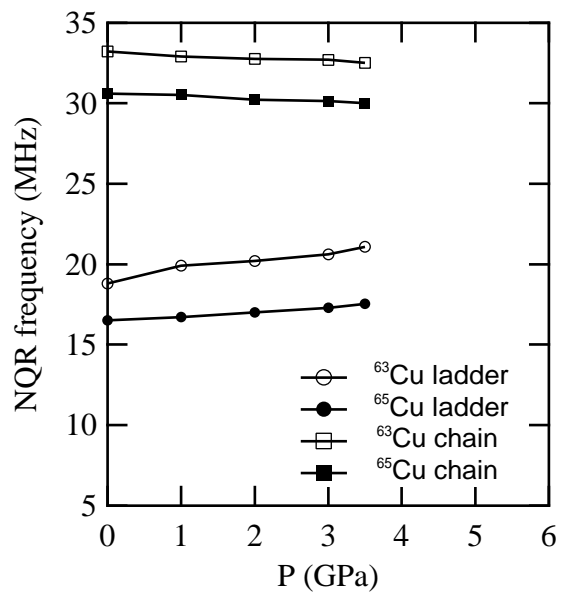}
\caption{\label{fig:epsart}Pressure dependence of the NQR frequency
of $^{63}$Cu and $^{65}$Cu nuclei.}
\end{figure}
The hole number is related to the orbital susceptibility
$\chi_{orb}$, namely $K_{orb}^c$. The orbital susceptibility for
$\emph{\textbf{H}}
\parallel$ c is expressed in the same way with high-\emph{$T_c$} cuprates $^{31}$ as
\begin{equation}
\frac{\chi_{orb}}{N}=n_{3d} \frac{2\mu_B^2}{E_{xy}-E_{x^2-y^2}}
\end{equation}
where N is the Avogadoro number and $\mu_B$ is the Bohr magnetron.
$E_{xy}$ and $E_{x^2-y^2}$ represent energy levels of the $3d_{xy}$
and $3d_{x^2-y^2}$ orbitals, respectively. The energy difference
between them is estimated to be 4.8 eV from the cluster model.
$^{32}$ \emph{$K_{orb}^c$} is linked with $\chi_{orb}$ as
\begin{equation}
K_{orb}^c= A^{orb}\chi_{orb}/N\mu_B
\end{equation}
where $A^{orb}(\equiv 2\mu_B<1/r^3>), <1/r^3>$ being a radius
average for a Cu ion, is a hyperfine field due to the orbital moment
induced by the applied field. $A^{orb}$ is estimated as 1028
(kOe/$\mu_B$) using $<1/r^3>=8.252$ $a.u.$, a value for a free
Cu$^{2+}$ ion. $^{33}$ From Eqs. (4) and (5), $K_{orb}^c=0.248n_{3d}
(\%)$ is obtained. A $K_{orb}^c$ of 0.18\% is estimated in a
pressure
 of 3.5-3.8 GPa. The estimate shows that $K_{spin}^c$ is enhanced at the optimum pressure.

The increase in $T_c$ or $D(E_F)$ is determined by two factors; the
hole number and the band width. The hole number seems to increase
beyond the optimal pressure, as is expected from Fig. 5, whereas the
band width tends to broaden with increasing pressure. Therefore, the
dome-shaped $T_c$ curve arises from naive balance between the two
factors: the transfer of holes to the ladder sites increases DOS
below the optimal pressure, whereas the broadening of the band width
would cause a decrease in DOS over the optimal pressure.

Anomalous behavior, namely that $K_{spin}^c$ is finite below $T_c$,
is actually observed at rather high fields. If it is related to
quasi one dimensionality, the singlet-FFLO phase transition is a
candidate that can explain the anomaly. In this scenario,
$K_{spin}^c$ or $\chi_{spin}$ at a low field would decrease with
decreasing temperature below $T_c$. Although NMR measurements at a
low field is important, it is difficult to curry out because the
overlapped NQR signals in Fig. 4 hardly split at a low field.

In summary, we have measured $1/T_1$ at an optimal pressure of 3.8
GPa, and observed transitional features from the spin-gapped state
to the Fermi-liquid state; the activated-type contribution fades
out, and approaches the Korringa relation with increasing pressure.
This implies that antiferromagnetic spin fluctuation is not
important unlike high-$T_c$ cuprates. The enhancement of $T_c$, or
an increase in DOS at the Fermi energy originates from an increase
in the number of holes transferred from the chain sites.

The authors wish to thank S. Fujimoto, and K. Kojima for their
fruitful discussions and K. Tatsumi and A. Hisada for their
experimental support. The present work was partially supported by
Grant-in-Aid for the Ministry of Education, Science and Culture,
Japan.





  \ \\

\end{document}